\documentstyle[referee,epsfig]{l-aa}

\begin{document}

   \thesaurus{20         
              (09.04.1;  
		09.13.2; 
		11.19.2; 
		11.09.1; 
               11.09.4;  
               13.09.1)  
              }
   \title{Dust properties of external galaxies; NGC 891 revisited}

   \subtitle{}

   \author{P.B.Alton\inst{1}
	\and E.M.Xilouris\inst{2}
	\and S.Bianchi\inst{1}
	\and J.Davies\inst{1}
	\and N.Kylafis\inst{2}
}

   \offprints{P.B.Alton, paul.alton@astro.cf.ac.uk}

   \institute {Department of Physics \& Astronomy, University of Wales, PO Box 913, Cardiff CF2 3YB, U.K.
        \and Foundation for Research \& Technology-Hellas, PO Box 1527, 711 10 Herakli
on, Crete, Greece
}

   \date{Received ??; accepted ??}

   \maketitle

   \begin{abstract}

We compare $850\mu$m SCUBA images of NGC 891 with the corresponding V-band optical depth predicted from radiation transfer simulations. These two tracers of dust show a very similar distribution along the minor axis and a reasonable agreement along the major axis. Assuming that the grains responsible for optical extinction are also the source of $850\mu$m emission we derive a submillimeter emissivity (emission efficiency) for dust in the NGC 891 disk. This quantity is found to be a factor of 2-3 higher than the generally-accepted (but highly uncertain) values adopted for the Milky Way. It should be stated, however, that if a substantial fraction of dust in NGC 891 is clumped, the emissivity in the two galaxies may be quite similar. We use our newly-acquired emissivity to convert our $850\mu$m images into detailed maps of dust mass and, utilizing 21cm and CO-emission data for NGC 891, derive the gas-to-dust ratio along the disk. We compute an average ratio of 260 -- a value consistent with the Milky Way and external spirals within the uncertainties in deriving both the dust mass and the quantity of molecular gas. The bulk of dust in NGC 891 appears to be closely associated with the molecular gas phase although it may start to follow the distribution of atomic hydrogen at radii $>$9 kpc (i.e.~$> \frac{1}{2}R_{25}$). Using the optical depth of the NGC 891 disk, we quantify how light emitted at high redshift is attenuated by dust residing in foreground spirals. For B-band observations of galaxies typically found in the Hubble Deep Field, the amount of light lost is expected to be small ($\sim$ 5\%). This value depends critically on the maximum radial extent of cold dust in spiral disks (which is poorly known). It may also represent a lower limit if galaxies expel dust over time into the intergalactic medium.

   \end{abstract}

%

\section{Introduction}
\label{intro}

The introduction of submillimeter (submm), imaging arrays such as SCUBA (Submillimeter Common User Bolometer Array; Holland et al 1999), heralds a revolution in our understanding of the dust properties of external spiral galaxies. For the first time, we can map cold (15-20K) interstellar grains with high sensitivity and good spatial resolution ($\sim 10''$) (Chini et al 1995; Hughes et al 1997; Clements et al 1993). Previous studies of radiation emitted by dust have relied heavily on measurements carried out by the Infrared Astronomical Satellite (IRAS) at 60 and $100\mu$m. These wavebands are sensitive primarily to warm ($\sim$ 30K) dust, and as such, are unlikely to give an accurate impression of either the true quantity or distribution of interstellar grains in external disks (Alton et al 1998a; Devereux \& Young 1990). Furthermore, satellite missions such as IRAS, and the recently completed Infrared Space Observatory (ISO) project, are constrained at present to using small mirrors (0.6-m c.f.~SCUBA 15-m dish). This limits their resolving capability to $\sim 1'$ in the far-infrared (FIR) which is only adequate for probing disk structure in the very closest spiral galaxies (Alton et al 1998b).\\

One of the first studies of nearby galaxies, using SCUBA, was carried out by Alton et al (1998c) who mapped the edge-on galaxy NGC 891 at 450 and $850\mu$m (see also Israel et al 1999). By carefully comparing the submm emission over the same region as the FIR flux ($60-100{\mu}m$), they found that this galaxy contains about an order of magnitude more dust than would have been inferred from IRAS observations alone. This same conclusion was reached by Xilouris et al (1999) who fitted the optical and near-infrared surface photometry of both NGC 891 and 6 other nearby, edge-on spirals with a sophisticated radiation transfer simulation. Their model, which takes account of both scattering and absorption by interstellar grains, only produces a self-consistent solution across several wavebands if the extinction lane harbours about 10 times more dust than might be expected from IRAS measurements.\\

In this paper, we relate the extinction modelling carried out for NGC 891 (Xilouris et al 1998; hereafter XAD) with the corresponding submm images acquired from SCUBA (Alton et al 1998c; hereafter ABR). In doing so, we address some important issues, namely:

\begin{enumerate}

\item whether the submm emission, which accounts for the bulk of grain material in NGC 891, correlates closely with the dust evident in the extinction lane (\S\ref{rtm});

\item in the case of a positive correlation, whether the submm emissivity in NGC 891 (or, equivalently, the grain mass-absorption coefficient) is comparable to that recorded for our own galaxy (\S\ref{emissivity});

\item given that we recover most of the dust mass from our submm images, how the gas-to-dust ratio varies along the major axis of NGC 891(\S\ref{gdratio});

\item how the submm emission is distributed with respect to the neutral gas phases (HI,H$_{2}$) and the stars in the disk (\S\ref{gdratio}). This information will prove crucial to the growing number of submm surveys, where sources are poorly resolved and subsidary observational data are limited.

\end{enumerate}

We begin by summarizing the salient results from XAD and ABR (\S\ref{submm} and \S\ref{rtm}) before seeking to relate the submm to the extinction dust lane.

\section{Submillimeter Maps}
\label{submm}

ABR obtained the deepest images yet of a nearby galaxy in the submillimeter waveband. They detected emission from $\frac{2}{3}$ of the optical disk, down to a noise-limited surface brightness of 3.5mJy/$16''$beam and 13mJy/$10''$beam at 850 and $450\mu$m respectively. At both submm wavelengths, the flux exhibits a pronounced peak at the nucleus and secondary maxima $\simeq 1'$ either side of the centre (equivalent to 3 kpc for a distance of 9.5 Mpc, adopted by both ABR and XAD, to NGC 891). The secondary maxima may be attributable to a ring structure seen edge-on or limb-brightening associated with spiral arms. We do not present the SCUBA maps here, but refer to ABR for both these images and more technical details {\em vis-a-vis} the observing procedure. The submm maps are much more radially-extended than the corresponding emission detected by IRAS, indicating that longer wavelength FIR radiation dominates at larger galactic radii. \\

ABR made a careful estimate of the ratio of cold dust to warm grains by smoothing all their submm/FIR images of NGC 891 (60,100,450 \& $850\mu$m) to the same resolution and then dividing the major axis into spatially-independent radial bins. This uniform treatment of the data was considered of major importance because there had been a tendency, in the past, to compare photometry at different wavelengths without regard for, or knowledge of, differences in the size of the relevant emitting region(s) (e.g.~Guelin et al 1993; Israel et al 1999; Clements et al 1993). The flux density emanating from each radial bin was matched by the superposition of 2 greybody curves. Thus:

\begin{equation}
\label{eq1}
F(\lambda) = \frac{\sigma}{D^{2}} Q({\lambda}) \Bigr( N_{W} B({\lambda},T_{W}) + N_{C} B({\lambda},T_{C}) \Bigr)
\end{equation}

\noindent where $F(\lambda)$ is the flux density at wavelength $\lambda$ (in Wm$^{-2}$Hz$^{-1}$), $\sigma$ is the geometrical cross-sectional area of an interstellar grain, $D$ is the distance to NGC 891 and $Q(\lambda)$ is the submm emissivity. $N_{W}$ and $N_{C}$ are the number of emitters (grains) in the warm and cold dust components respectively whilst $B(T_{W})$ and $B(T_{C})$ represent the corresponding blackbody intensities. $Q(\lambda)$ is sometimes refered to as the emission efficiency and can be considered as the ratio of the emission cross-section to the geometrical cross-section of the grain (Spitzer 1978; Whittet 1992). \\

ABR found that all radial bins could be fitted well with $T_{C}$=15-20K \footnote{in the original paper, $T_{C}$ was given as $15\pm$6K but the more correct range is 15-20K} and $T_{W}$=30-37K, and $\frac{N_{C}}{N_{W}}$ increased from about 26 at the nucleus to 54 at $\simeq \frac{1}{2} R_{25}$. The {\em ratio} $\frac{N_{C}}{N_{W}}$ is fairly secure. {\em However}, the exact dust masses pertaining to $N_{C}$ and $N_{W}$ depend critically on the absolute level of $Q(\lambda)$ which, as we shall see, is highly uncertain. At this stage, it is important to emphasize that over 90\% of the $850\mu$m  emission arises, at all radii, from the cold (15-20K) dust component and that this material, in turn, constitutes the bulk of galactic dust. In other words, our $850\mu$m SCUBA map indicates, almost completely, how galactic dust is distributed within NGC 891. Moreover, if we could avail ourselves of a well-determined value for the grain emissivity at $850\mu$m, $Q({850\mu}m)$, our SCUBA image would yield, to a good approximation, the total dust mass in the disk.\\

The submm/FIR emissivity, $Q(\lambda)$, is a poorly known quantity and virtually the only sources of information stem from Hildebrand (1983) and Draine \& Lee (1984), which are cited almost exclusively in the literature. The measurement of Hildebrand (1983) is based on Kuiper observations (55 and $125\mu$m) of a single Galactic reflection nebula, NGC 7023, where attenuation of ultraviolet light from the central star is effectively used to constrain the amount of dust present. In order to account for the observed FIR flux densities, the emissivity at $125\mu$m was estimated to lie in the range $\frac{1}{230}-\frac{1}{1700}$. Although this represents little better than an order of magnitude determination, the emissivity of Hildebrand has been applied by FIR observers almost religiously since its derivation nearly 2 decades ago. This is at least partially attributable to a dearth of measurements in the interim years, in particular observations applicable to the diffuse interstellar medium (ISM) where grains may well differ from their particulate counterparts in reflection nebulae (Pendleton et al 1990; Mathis 1990). We note that Casey (1991) has repeated the Hildebrand procedure for 5 Galactic reflection nebulae (and derived an emissivity approaching the upper limit of the Hildebrand determination), but this work has been seldom cited in the literature. Draine and Lee (1984) have used a {\em pot pourri} of laboratory measurements and astronomical data to construct dielectric functions appropriate to graphite and silicate interstellar grains. In this way, they have derived a FIR emissivity close to the mean value given by Hildebrand. Finally, we note that Bianchi et al (1999) recently determined an emissivity value of $\frac{1}{500}$ for the diffuse dust in the Milky Way at $100\mu$m. Their technique relied on the tight spatial correlation observed between interstellar extinction and the FIR radiation detected by COBE. This allowed them, in a similar fashion to Hildebrand, to constrain the amount of dust responsible for the FIR emission. Depending on the exact behaviour of $Q$ with wavelength (discussed below), the Bianchi et al value is somewhere towards the higher end of the Hildebrand range.\\

In the following sections, we attempt an independent determination of the submm/FIR emissivity in NGC 891 by using extinction observed in the disk to constrain the amount of grain material present. This approach differs from previous treatments of extragalactic submm/mm observations where, to one extent or another, workers have tended to make use of dust properties in the Milky Way in order to infer quantities in external galaxies. For example, Guelin et al (1993) use a Galactic emissivity to turn their 1.3mm observations into a dust column density. Assuming a solar gas-to-dust ratio they then quantify how much (molecular) gas is present in the system. In a slightly different approach, Neininger et al (1996) use 21cm emission to infer a dust column density (again via a Galactic gas-to-dust ratio) and then compare this quantity with their observed submm/mm emission in order to derive the corresponding emissivity. Our current method is based on the technique pioneered by Hildebrand (1983). In essence, the dust column density is measured by means of optical/ultraviolet extinction and this quantity is then compared to the observed FIR/submm emission in order to infer the long wavelength emissivity. In deriving $Q(\lambda)$ we make no direct use of Galactic properties as such. Whilst we follow the same general method as Hildebrand, we believe that the submm window produces more robust values of emissivity than fits made to the emission near the greybody peak (Hildebrand fits near $100\mu$m). The reason for this is that, for any observed level of FIR/submm flux, the implied dust mass will vary at least as strongly as $T^{4}$ for emission near the peak but only $T^{-1}$ in the Rayleigh-Jean tail (where $T$ is the grain temperature). If the dust mass is fixed {\em a priori} by the amount of extinction, the corresponding uncertainties are then transfered to the derived values of emissivity. Thus $Q(\lambda=100{\mu}m)$ will be much more dependent on the grain temperature than $Q(\lambda=850{\mu}m)$ and, as a consequence, far more uncertain (see Hughes et al 1997 for an indepth discussion).\\

Before we are able to embark on this process for NGC 891, we have to be sure that the submm, which traces the bulk of the dust in this galaxy, corresponds closely to the grain material responsible for the absorption lane. To this end, we briefly describe the photometric fit made to NGC 891 by XAD which yields directly the optical depth at each point in the disk.

\section{Radiation Transfer Model}
\label{rtm}

The approach adopted by XAD (see also Xilouris et al 1997;1999) is to simulate the optical/NIR appearance of edge-on spirals by creating a model galaxy consisting of an exponential stellar disk, a R$^{\frac{1}{4}}$ bulge and an exponential dust distribution. A pixel-to-pixel comparison is made between the real and simulated object in order to fit scale-heights (z-direction) and scale-lengths (radial direction) to both the stellar and dust disks and to determine the optical depth through the centre of the galaxy (as seen face-on). In addition to radiation absorption, the model takes account of photon scattering in the disk by adopting the Henyey-Greenstein phase function (Henyey \& Greenstein 1941; Bohren \& Huffman 1983). Our confidence in the model is significantly boosted by the fact that the output parameters are highly consistent across several optical and NIR wavebands. Thus for NGC 891, the radial scale-length for the grain distribution is calculated to be 8.1 kpc in the V-band whilst a determination in the remaining filters (K,J,I,B) strays by only 5-10\% from this value. Simulations for 7 edge-on galaxies indicate, in all cases, an extinction lane consistent with Galactic-type reddening. It is important to emphasize that our modelling of the dust lane vastly transcends simplistic screen models (which have often been applied in the past) and takes full account of mixing between stars and dust. Admittedly, one obvious limitation to our simulation is that no account is taken of dust clumping or indeed spiral structure within the disk. Kylafis et al (1999) have tested the influence of spiral structure on the radiation transfer in NGC 891 and found that it produces almost no change to the large-scale properties inferred from the model. Clumping, on the other hand, may well `hide' large amounts of dust so that it cannot be detected by a fit to the large-scale extinction. We postpone a discussion on the effects that clumping might have on our results until later (\S\ref{discussion}).\\

\begin{figure}[hp!]
\centering
\resizebox{!}{21cm}{\includegraphics*[0mm,35mm][210mm,280mm]{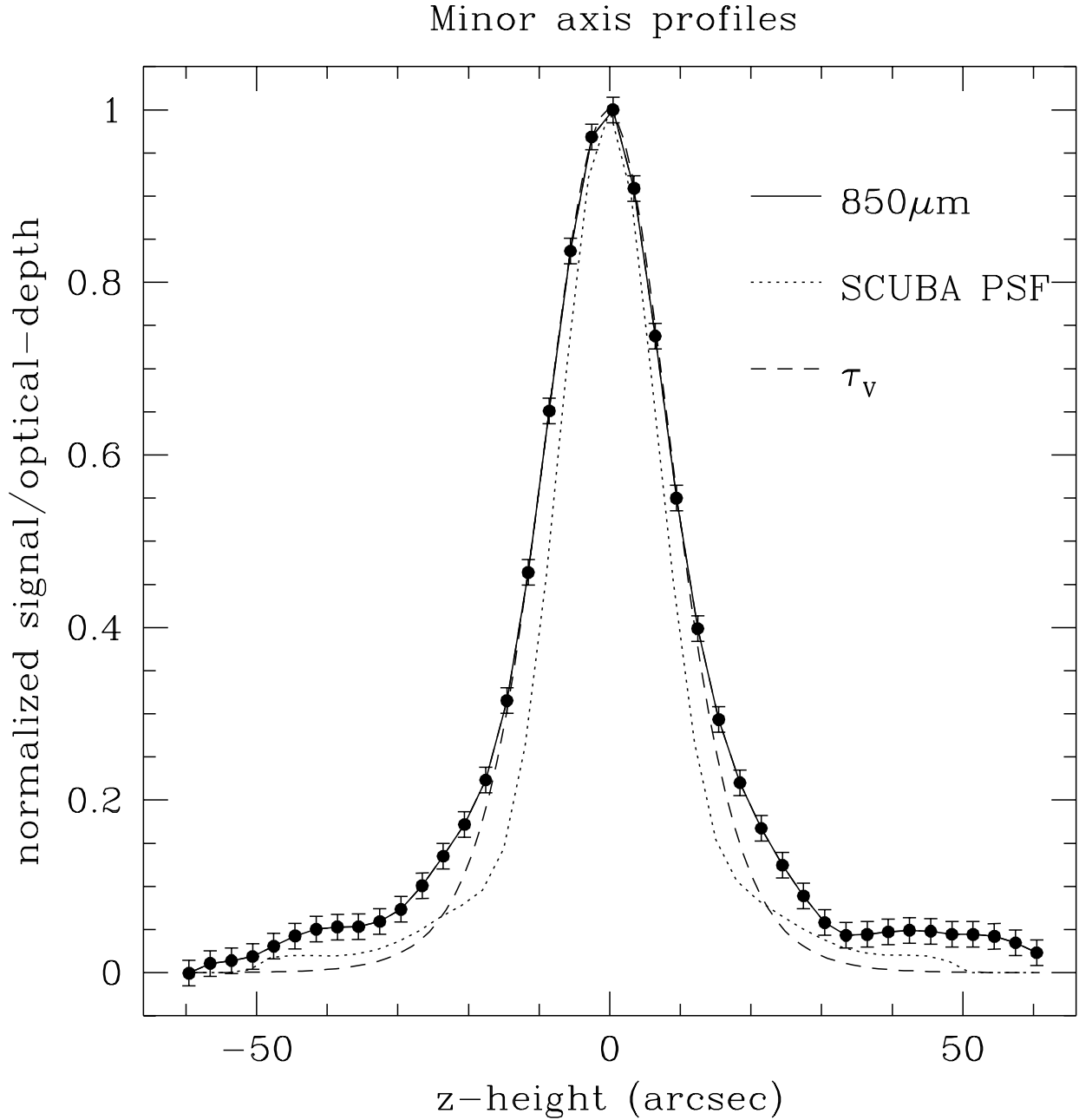}}
\caption{Submm profile along the minor axis of NGC 891 compared with the corresponding extinction model of Xilouris et al (1998). The solid curve and markers denote the average profile at $850\mu$m measured by SCUBA. The dashed line represents the V-band optical depth ($\tau_{V}$) that is predicted from radiation transfer modelling, after convolving to the same spatial resolution as the submm data. The dotted line shows the point spread function (PSF) measured for the submm data. All curves have been normalized to a value of unity at zero z-height.}
\label{fig2}
\end{figure}

\begin{figure}[hp!]
\centering
\resizebox{!}{21cm}{\includegraphics*[0mm,35mm][210mm,280mm]{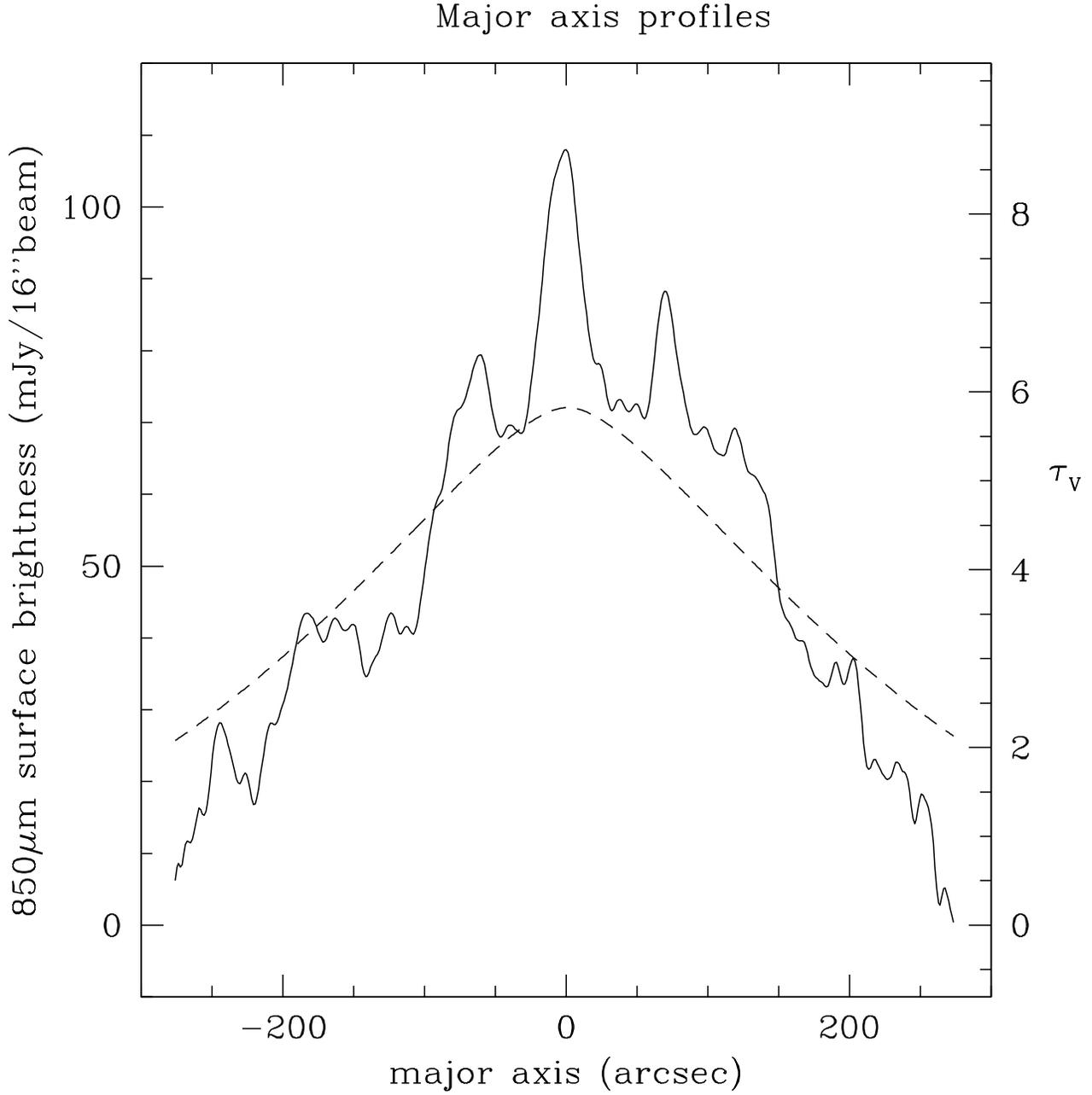}}
\caption{Submm profile along the major axis of NGC 891 (solid line) compared with the corresponding extinction model of Xilouris et al 1999 (dashed line). The profile in optical depth ($\tau_{V}$) was created only after the extinction model had been convolved to the same spatial resolution as the submm data (this explains why $\tau_{V}$ here is a factor 4 or so lower than the opacity derived directly from optical images). Positive distances along the major axis correspond to the northeast half of the disk. $1\sigma$ photon errors in the $850\mu$m profile are 2 mJy/$16''$beam.}
\label{fig3}
\end{figure}

Since we shall be concentrating on the V-band when making a comparison between extinction in NGC 891 and the submm emission detected by ABR, we reproduce in table~\ref{tab1} the relevant parameters fitted by XAD at this wavelength. Using the tabulated parameters, we can construct a synthetic map of V-band optical depth ($\tau_{V}$), which, after smoothing to the same spatial resolution as the $850\mu$m SCUBA map ($16''$ FWHM), can be compared directly with the submm image. In figures \ref{fig2} and \ref{fig3}, we display profiles of $\tau_{V}$ along the minor and major axes of the galaxy model after the smoothing process. The corresponding $850\mu$m cross-sections are also shown. In all cases, a width of $40''$ ($2.5\times$ beam FWHM) is used to sample the emission/optical-depth perpendicular to the profile direction. The submm profile, in fig.~\ref{fig2}, is the mean of several transects across the major axis taken at various distances from the galactic nucleus. The average profile has been normalized, along with the $\tau_{V}$ curve, to unity at zero z-height. We can see that the observed and model profiles are extremely similar in this direction (a deconvolved size of $13{\pm}1''$ or 610 pc for both FWHM). To some extent, this similarity may be attributable to the point spread function (PSF) being quite large with respect to the intrinsic dust layer. However, by artificially varying the scale-height of $\tau_{V}$, before smoothing, we have assured ourselves that the thickness of the absorption lane must lie within 30\% of the submm width. For z-heights of $20-50''$ (900-2300 pc), there is an excess at $850\mu$m over what would be expected purely from the extinction model. The SCUBA beam is principally gaussian but slight wings in the PSF may contribute to this observed excess. In fact, until a more careful comparison is made between the high-z `emission' and the PSF wings (Alton et al 1999a), we are not in a position to attribute this excess to the `vertical' dust chimneys evident in optical images of NGC 891 (Howk \& Savage 1997; Dettmar 1990).\\

For profiles along the major axis, we plot the optical depth given by the model against the observed $850\mu$m surface brightness (fig.~\ref{fig3}). It should be noted that when the comparison is made here, it is after the opacity model has been smoothed to the same resolution as the submm image (thus $\tau_{V}$ is a factor of 4 lower than the optical depth directly inferred from the original optical image). The correspondance along the major axis is not as good as that for the minor axis and can only be described as fair. Due to its simplicity, we clearly cannot expect the XAD model to reproduce the local fluctuations in the submm profile. However, on average, the $\tau_{V}$ curve still appears to be somewhat more extended than the $850\mu$m emission. In fact, if we could choose a $\tau_{V}$ curve that would match the behaviour of SCUBA data perfectly, the dust model would have a scale-length of $\simeq$5.3 kpc as opposed to 8.1 kpc derived by XAD. Under such circumstances the dust layer would possess the same radial fall-off as the V-band stars. The ratio between optical depth and $850\mu$m surface brightness, averaged over the profiles in fig.~\ref{fig3}, is as follows:

\begin{equation}
\label{eq2}
\frac{\tau_{V}}{f_{850}} = 81 \pm 22
\end{equation}

\noindent where $f_{850}$ is the submm surface brightness, in Jy/$16''$beam, and the uncertainty represents the standard deviation of a linear fit between $f_{850}$ and $\tau_{V}$. \\

Although the major axis fit between submm emission and optical attenuation is only fair, we will assume that in both cases the same population of grains manifests itself. Certainly, simulations of dust bathed in interstellar radiation fields suggest that the classical `big grains', which are responsible for extinction (size $\sim 0.1{\mu}$m), increasingly dominate the emission process at wavelengths beyond $60\mu$m (Desert et al 1990). And indeed, observationally, Bianchi et al (1998) have recently established a convincing correlation between optical extinction and submm emission in the nearby spiral NGC 7331. For NGC 891, we therefore fix the $850\mu$m surface brightness with respect to the visual optical depth, $\tau_{V}$, according to eq.~\ref{eq1}. By doing this, we will be able to derive the dust emissivity at $850\mu$m. This calculation is carried out in the next section. 

\section{Submillimeter Emissivity}
\label{emissivity}
\subsection{Derivation}
\label{derivation}

As mentioned in \S\ref{submm}, the $850\mu$m waveband is completely dominated by cold dust (15-20K) at all positions in the disk. In fact, the Rayleigh-Jeans tail of the warm grain component contributes about 8\% to the flux density in this filter. It is possible, therefore, to rewrite eq.~\ref{eq1} approximately as follows:

\begin{equation}
\label{eq3}
0.92 \times F(850{\mu}m) = \frac{N_{C} \sigma}{D^{2}} Q(850{\mu}m) B(850{\mu}m,15-20K)
\end{equation}

\noindent where $N_{C}$, the number of cold grains, comprises the bulk of interstellar dust in NGC 891. In order to be able to use the ratio $\frac{\tau_{V}}{f_{850}}$ (eq.~\ref{eq2}) in this relation, we have to refer to the definition of optical depth, namely:

\begin{equation}
\label{eq4}
\tau_{V} = N_{d} \sigma Q(V)
\end{equation}

\noindent where $N_{d}$ is the dust column density (grains per unit surface area of galaxy), $\sigma$, as before, represents the geometrical grain cross-section and $Q(V)$ constitutes the emissivity, or extinction efficiency, in the V-band.\\

\noindent If we can derive the flux density per unit surface area, $N_{C}$ in eq.~\ref{eq3} becomes $N_{d}$ and we will be in a position to substitute $\tau_{V}$, from eq.~\ref{eq4}, into eq.~\ref{eq3}. To do this we use $f_{850}$, the $850\mu$m surface brightness (in Jy/$16''$beam). Dividing $f_{850}$ by the beam area (in m$^{2}$), and multiplying by $10^{-26}$ to convert from Jansky to Wm$^{-2}$Hz$^{-1}$, we obtain the flux density per unit surface area (as required). This can be substituted into the left-hand side of eq.~\ref{eq3} thus:

\begin{equation}
\label{eq5}
0.92 \times \frac{f_{850} \times 10^{-26}}{4.55\times 10^{-9} D^{2}} = \frac{N_{d} \sigma}{D^{2}} Q(850{\mu}m) B(850{\mu}m,T=15-20K)
\end{equation}

\noindent where $N_{d}$ has now replaced $N_{C}$.\\

\noindent Substituting $N_{d}$ with $\frac{\tau_{V}}{\sigma Q(V)}$ (eq.~\ref{eq4}), we can eliminate the geometrical grain cross-section $\sigma$ (the distance D also falls out), leaving:

\begin{equation}
\label{eq6}
\frac{\tau_{V}}{f_{850}} = \frac{Q(V)}{Q(850{\mu}m)} \frac{2.02 \times 10^{-18}}{B(850{\mu}m,T=15-20K)}
\end{equation}

\noindent For the blackbody intensity, $B(850{\mu}m,T=15-20K)$, we adopt a grain temperature of $T=17.5$ K. Then, comparing with eq.~\ref{eq2}, we obtain the emissivity at $850\mu$m with respect to the V-band emissivity as:

\begin{equation}
\label{eq7}
\frac{Q(V)}{Q(850{\mu}m)} = 15900 \pm 4330
\end{equation}

\noindent To derive $Q(850{\mu}m)$ directly from eq.~\ref{eq7}, we must have some knowledge of $Q(V)$. For `classical' grains of radius $0.1\mu$m, which are responsible for optical extinction in the diffuse ISM of the Milky Way, $Q(V)$ lies between 1 and 2 (Whittet 1992; Alton 1996; Spitzer 1978). The interstellar grains of NGC 891 are unlikely to be much different in this respect, given that extinction within the disk corresponds extremely well to a Galactic reddening law (\S\ref{rtm}). Thus we use $Q(V)=1.5$ in eq.~\ref{eq7} to infer:

\begin{equation}
\label{eq8}
Q(850{\mu}m) = \frac{1}{10600\pm3000}
\end{equation}

\subsection{Discussion}
\label{discussion} 

\begin{figure}[hp!]
\centering
\resizebox{!}{21cm}{\includegraphics*[0mm,35mm][210mm,280mm]{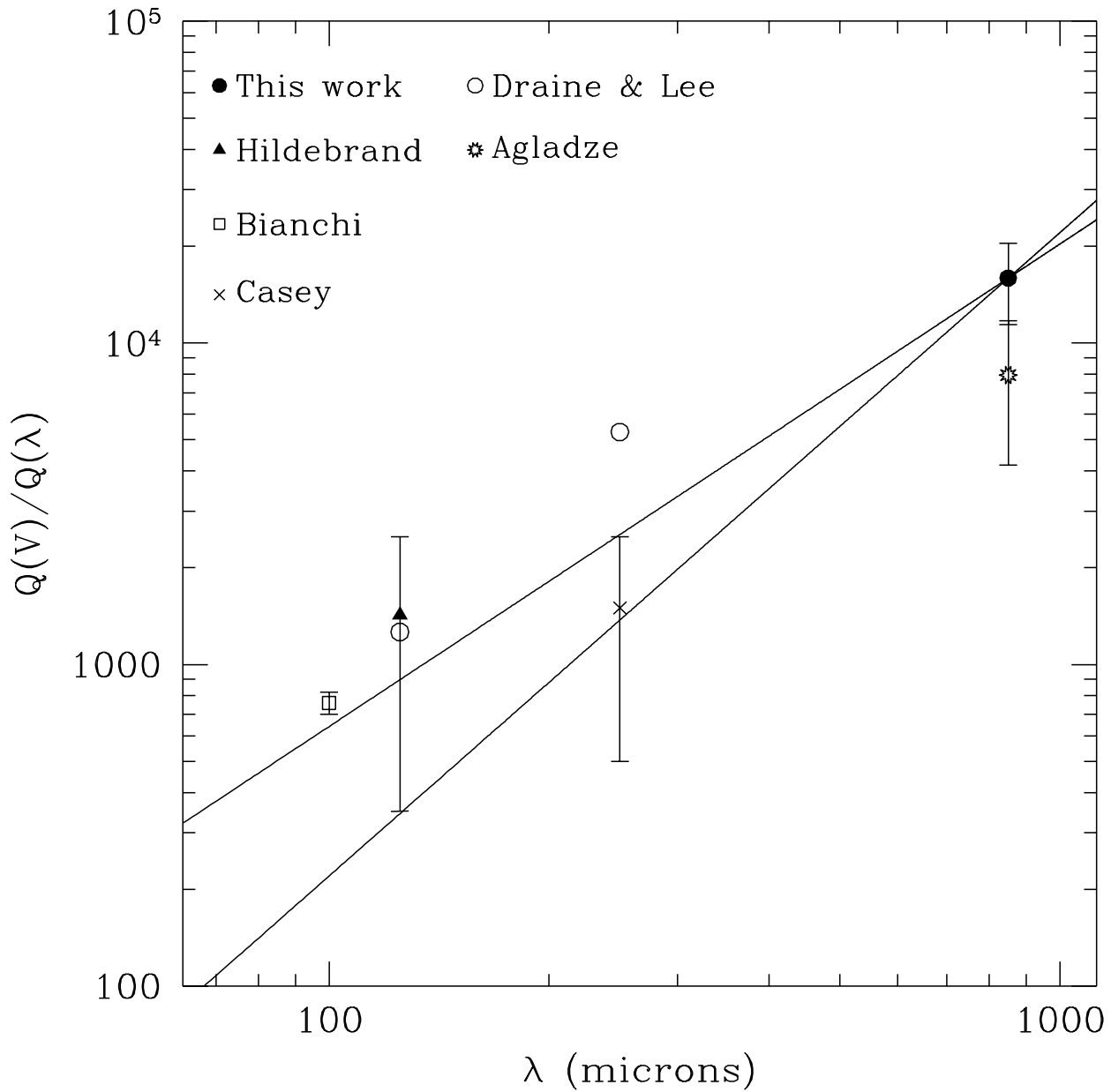}}
\caption{The FIR grain emissivity in NGC 891 compared to the Milky Way. The solid circle denotes the ratio $\frac{Q(V)}{Q(850{\mu}m)}$ evaluated in this work for NGC 891. Symbols at shorter wavelengths represent measurements for dust in the Galaxy. The lines passing through the NGC 891 data-point assume $Q(\lambda)$ scales as ${\lambda}^{-1.5}$ (top line) or ${\lambda}^{-2.0}$ (bottom line). The error bars for both NGC 891 (this work) and Bianchi et al (1999) represent the standard deviation in the linear correlation fitted between FIR/submm emission and optical depth $\tau_{V}$ (see text for details). The data point for Agladze et al (1994) assumes that astronomical grains consist of an olivine-like material whilst the corresponding error bar allows for the grains being amorphous (smaller $\frac{Q(V)}{Q(850{\mu}m)}$) or entirely crystalline (higher $\frac{Q(V)}{Q(850{\mu}m)}$).}
\label{fig4}
\end{figure}

The grain emissivity within the Milky Way has usually been estimated at shorter FIR wavelengths (e.g.~Hildebrand finds $\frac{Q(V)}{Q(125{\mu}m)} = 350-2500$). It is difficult, therefore, to compare Galactic data with eq.~\ref{eq8} unless we assume some sort of wavelength behaviour for the dust emissivity. Fortunately, a large amount of recent observational evidence seems to be converging on an emissivity, {\em for the general ISM of non-active spirals}, which varies as $\lambda^{-(\beta)}$ where the spectral index ${\beta}=1.5-2.0$ for $\lambda>100{\mu}$m. Indeed, rigorous FIR spectral measurements of Milky Way dust, using both COBE and balloon-borne instruments, imply a spectral index between 1.5 and 2.0 for the diffuse ISM in the Galaxy (Masi et al 1995; Reach et al 1995). Likewise, the relatively high flux density at $450\mu$m compared to emission at  $850\mu$m, for galaxies such as NGC 891 and NGC 7331, is indicative of an index $\geq 1.5$ (Bianchi et al 1998). Accordingly, we plot in fig.~\ref{fig4} the dust emissivity for NGC 891 (as a ratio $\frac{Q(V)}{Q(850{\mu}m)}$) assuming a wavelength dependency $\lambda^{-\beta}$ ($\beta=1.5-2$). Estimates for the FIR emissivity of Milky Way dust, as derived by Hildebrand (1983), Casey (1991), Draine \& Lee (1984) and Bianchi et al (1999b) (see \S\ref{submm}), are annotated on this plot. We also add the emissivity measured by Agladze et al (1994) for `astronomical-type' silicates tested at 20K under laboratory-controlled conditions.\\

The FIR grain emissivity in NGC 891 appears to be somewhat higher than most estimates of the same property in the Milky Way. In particular, the `canonical' values of Draine \& Lee (1984) and Hildebrand (1983) are about a factor of 2-3 lower than the quantity we have derived here. The discrepancy is admittedly smaller for an emissivity law following $\lambda^{-1.5}$, but both the Draine and Lee model and the Bianchi determination are consistent with $\beta=2$ and should, therefore, strictly speaking, be compared with the steepest line in fig.~\ref{fig4}. Similarly, the lowest values of $Q(V)/Q({850\mu}m)$, as given by Hildebrand, are derived under the assumption that $\beta=2$. Therefore, the agreement in this case can only be described as marginal. In this respect, the NGC 891 emissivity seems more consistent with the Casey (1991) observations (although why there should be such a difference between the Hildebrand (1983) and Casey (1991) measurements, when both apply to Galactic reflection nebulae, is not altogether clear). Although the FIR emissivity we predict for NGC 891 is somewhat higher than most values used for the Galaxy, it is still much lower than the inference from laboratory-controlled experiments. Indeed, the values of $Q(850{\mu}m)$, implied by Agladze et al, are astonishingly high, although the authors themselves do not offer an explanation as to why there is such an incongruence with astronomical observations. \\

It would certainly be something of a surprise if interstellar grains in NGC 891 possess a submm emissivity grossly different from Milky Way dust (particularly since XAD find that extinction within NGC 891 is characterized by a Galactic reddening law). The velocity fields and distribution of neutral gas within the disks of both systems has been noted by several authors as being strikingly similar (Guelin et al 1993; Garcia-Burillo et al 1992; Scoville et al 1993). However, it is possible that NGC 891 may display a somewhat more active halo with `thick disks' of both neutral and ionized gas extending to several kpc above the midplane (Dettmar 1990). Since this extended disk only contains a few percent of the total dust in NGC 891 (Howk \& Savage 1997; Alton et al 1999a), the activity at the disk-halo interface should not necessarily imply any general pecularities for the dust properties of the {\em main} disk. Although the error bar in fig.~\ref{fig4} represents only a 30\% standard deviation in the fit between the submm and opacity measurements, the total uncertainty in our estimate of $Q(850{\mu}m)$ is quite likely to be about a factor of 2 (due to $\sim33$\% error in both $Q_{V}$ and $B(850{\mu}m,T)$). Moreover, measurements of FIR emissivity for Galactic grains are likely to be uncertain by an order of magnitude once extrapolated to submm wavelengths (Hughes et al 1993). Therefore we should not be unduly worried by the factor 2-3 difference between the emissivity derived for NGC 891 and estimates of the corresponding property for the Milky Way. It is interesting to note that grain models which successfully reproduce the extinction law within our own Galaxy, can display radically different submm emissivities. For example, the fluffy agglomerates of silicate and carbon, that have been postulated by Matthis \& Whiffen (1989), satisfy the near-infrared to ultraviolet reddening law in much the same way as the Draine \& Lee (1984) model. However in comparison to the latter, they possess an emissivity nearly 3 times larger at $\lambda >300{\mu}m$.\\

As a confirmatory {\em addendum} to the preceding paragraph, we note that Dunne et al (1999) estimate the submm mass-absorption coefficient in NGC 891 ($\kappa_{850{\mu}m}$) by using our reported values of $850\mu$m flux density (ABR) and fixing the dust mass via depletion arguments rather than observed extinction (a constant fraction of interstellar metals are assumed to be bound up in grains). The Dunne et al analysis yields $\kappa_{850{\mu}m} = 0.24$ m$^{2}$kg$^{-1}$ which, for grains of radius $0.1\mu$m and material density 3000 kgm$^{-3}$, corresponds to $Q(850{\mu}m) = \frac{1}{10400}$, i.e. almost exactly the value we derive in eq.~\ref{eq8}.\footnote{the mass absorption coefficient and grain emissivity are related in the following way: $\kappa_{850{\mu}m} = \frac{3Q({850\mu}m)}{4a{\rho}}$ }\\

As mentioned previously, one cause for concern in our derivation of submm emissivity is the fact that the opacity simulations of XAD do not account for dust clumping. This will tend to {\em underestimate} the amount of dust causing extinction within the disk. Witt et al (1999; also Witt priv.~comm.) have analysed V-K excesses from spirals disks using homogeneous and clumpy-phase models and found that the latter typically require 50\% more dust mass for the same level of extinction. Similarly, the Monte Carlo radiative transfer simulations of Bianchi et al (1999b), which assign dust clumps according to the distribution of molecular gas, indicate that homogeneous models, such as XAD, will underestimate the dust mass by a factor of 2 for spiral galaxies viewed edge-on. We have attempted to assess the effects of clumping in NGC 891 ourselves, by estimating the amount of dust in gas clouds too optically thick to be detected satisfactorily by the XAD homogeneous model. Our reddest waveband is K, therefore our technique becomes insensitive to environments with $A_{K}>1$ ($A_{V}>11$). Within the Galaxy, this corresponds to a column density $N(H)$ of $\sim 2\times10^{22}$ cm$^{-2}$ (Bohlin et al 1978) which is close to the column density within most giant molecular clouds (GMCs) [see for example Larson (1981), who found $N(H) = 2.1\times10^{22}$ cm$^{-2}$ / $L^{0.1}$ for cloud sizes of $L$=0.05-100pc]. It is possible, then, that we may `miss' a sizeable fraction of grains residing in GMCs. By comparing relative emission strengths detected from $^{12}$CO and $^{13}$CO molecules in various gas clouds, Polk et al (1988) estimate that 50\% of molecular gas resides in Galactic GMCs and the remainder constitutes a more diffuse medium surrounding GMCs. This information, then, suggests that the XAD model will `overlook' about 50\% of the H$_{2}$ phase and, in total, perhaps 30\% of grains associated with the (HI+H$_{2}$) gas phase of NGC 891. Since the submm is optically thin, any increase in the amount of dust inferred from extinction would decrease $Q(850{\mu}m)$ in eq.~\ref{eq8}. Indeed, a 50-100\% rise in $\tau_{V}$, due to clumping, will yield a value of $\frac{1}{16000}$ -- $\frac{1}{21000}$ for NGC 891. This is much closer to extrapolations of the Galactic emissivity into the submm waveband ($Q(850{\mu}m) \sim \frac{1}{27000}$). Intriguingly, if dust clumps are preferentially located towards the centre of NGC 891 (where the molecular gas is known to be concentrated; Guelin et al 1993), a correction for clumping would steepen the $\tau_{V}$ profile in fig.~\ref{fig3} to a shape more akin to the submm distribution.\\

The submm emissivity we have derived for NGC 891 (eq.~\ref{eq8}) can now be used, in conjunction with our SCUBA images, to create dust maps for the disk. In the following section, these maps are employed to derive the gas-to-dust mass ratio along the major axis. Before embarking on this part of the analysis, however, we recap the method we have used to find dust masses in NGC 891. In lieu of extrapolating the FIR emissivity estimated for the Milky Way to submm observations of NGC 891, we have taken the visual optical depth calculated from radiation transfer modelling (XAD), and effectively constrained the number of grains, $N_{C}$, appearing in eq.~\ref{eq3}. This technique relies on knowledge of the extinction efficiency in the V-band, $Q(V)$. However, this quantity is relatively well known ($1.5\pm 0.5$), compared with the near order of magnitude uncertainty in $Q(850{\mu}m)$ assumed for the Milky Way (Hughes et al 1993; Hughes et al 1997). Fixing $N_{C}$, allows us to use eq.~\ref{eq3} to derive the $850\mu$m emissivity from the $850\mu$m flux densities in our SCUBA maps.\\

Now that we have a robust estimate of $Q(850{\mu}m)$, we use the following formula to derive the dust mass, $M_{d}$ (kg), at each point in the disk:

\begin{equation}
\label{eq9}
M_{d} = \frac{4 a \rho D^{2}}{3} \frac{0.92 \times F(850{\mu}m)}{Q(850{\mu}m) B(850{\mu}m,15-20K)}
\end{equation}

\noindent where $a$ and $\rho$ are the radius and material density of the interstellar grains, respectively, and $D$ is the distance to NGC 891 (see, for example, Hildebrand (1983) for derivation of eq.~\ref{eq9}). The determination of $M_{d}$ {\em will inevitably} rely on assumptions about the grain properties (typically $a=0.1\mu$m and $\rho=3\times 10^{3}$ kgm$^{-3}$ are applied), and, furthermore, some uncertainty is introduced by $D$ appearing as the square in eq.~\ref{eq9}. However, the evaluation of $M_{d}$ is greatly improved by our newly-acquired knowledge of $Q(850{\mu}m)$. Moreover, when we consider the neutral gas in NGC 891, we can expect the gas-to-dust mass ratio to be independent of distance. As a final note, we underline the fact that the blackbody intensity in eq.~\ref{eq9}, $B(850{\mu}m,15-20K)$, varies linearly with temperature because we are in the Rayleigh-Jeans tail of the greybody curve. This means that $B(850{\mu}m,15-20K)$ will change by as little as $\frac{1}{3}$ over the suggested temperature range of 15-20K.\\

It is important to point out that, whilst we use eq.~\ref{eq9} to infer the dust distribution from our $850\mu$m map of NGC 891, the overall dust mass has already been normalized to $\tau_{V}$. This is because we have substituted $\tau_{V}$ into eqs.~\ref{eq6}-\ref{eq8} in order to determine $Q(850{\mu}m)$. Thus, whilst our SCUBA map provides the detailed spatial information concerning the dust distribution, the total amount of dust will not be any different from what we would have directly inferred from the optical depth model of XAD.

\section{Gas-to-dust Mass Ratio}
\label{gdratio}

We produce a dust map of NGC 891 by using D=9.5 Mpc, $a=0.1\mu$m and $\rho=3\times 10^{3}$ kgm$^{-3}$ in equation \ref{eq9}. The grain distribution can then also be expressed as:

\begin{equation}
\label{eq10}
M_{d}/16''beam = 8.52({\pm}2.31) \times 10^{36} f_{850}
\end{equation}

\noindent where $f_{850}$ is the $850\mu$m surface brightness in Jy/$16''$beam.\\

Due to the proximity of NGC 891 the neutral gas in this galaxy has been reasonably well investigated. Rupen et al (1991) mapped out the atomic hydrogen with a high sensitivity and an effective beam of $20''$. The atomic gas distribution can be described as a `plateau' which extends approximately out to the $R_{25}$ but is sharply truncated at this radius. Within the inner few kpc of the stellar disk there is a relative dearth of HI. The molecular gas, on the other hand, appears to be fairly centrally concentrated, with a distribution similar to the submm emission detected by SCUBA. Scoville et al (1993) employed high resolution interferometry ($2.3''$) to map NGC 891 out to radius of 10 kpc in the CO (J=1-0) line. Emission appears to be restricted to a thin disk ($\sim 4''$ FWHM) with a radial distribution reminiscent of our own Galaxy (Young \& Scoville 1991).\\

In deriving the gas-to-dust mass ratio in NGC 891 we make use of these neutral gas measurements but ensure that all the relevant data are smoothed to a common spatial resolution.\footnote{the integrated flux in the Scoville et al interferometric observations agrees within 5\% with the single disk map of Garcia-Burillo et al (1992) assuring us that we are not missing any sizeable large-scale emission.} Thus, we took the CO interferometry data of Scoville et al and the dust map based on the $850\mu$m observations, and convolved them with a gaussian of $19.9''$ and $12.4''$ FWHM respectively. This produced images of comparable resolution to the Rupen et al map at 21cm. We then profiled along the major axis of each image using a bin width of $40''$ and a sampling interval of $8.4''$. To convert the CO emission to a H$_{2}$ column density we used a value of $2 \times 10^{20}$ cm$^{-2}$ K km/s for the conversion factor $X$. There is considerable controversy surrounding the exact value of $X$ (estimates vary from about 1.5-6 $\times 10^{20}$ cm$^{-2}$ K km/s for quiescent, giant spiral disks) but $1.8 \times 10^{20}$ seems appropriate for the general ISM in the Milky Way (Maloney 1990). Ultimately, we wish to compare the gas-to-dust ratio in NGC 891 with that prevailing in the Galaxy so, in lieu of any direct measurement of NGC 891, we adopt $X=2\times 10^{20}$ cm$^{-2}$ K km/s. At this stage, we refrain from allowing $X$ to vary with radius (as might be the case for a steep metallicity gradient between the centre and edge of the NGC 891 disk). Maloney (1990) provides a cogent argument as to why corrections to $X$ may only be necessary for environments with a metallicity {\em below} solar. Virtually nothing is known about the metallicity in NGC 891 (its edge-on orientation precludes optical spectroscopy) but, so long as NGC 891 constitutes a typical giant spiral galaxy, the heavy element abundance is unlikely to be lower than solar for those parts of the disk where Scoville et al detect CO emission (Garnet 1998). We check our assumptions about the metallicity in NGC 891 in the next section (\S\ref{metallicity}).\\

\begin{figure}[hp!]
\centering
\resizebox{!}{21cm}{\includegraphics*[0mm,35mm][210mm,280mm]{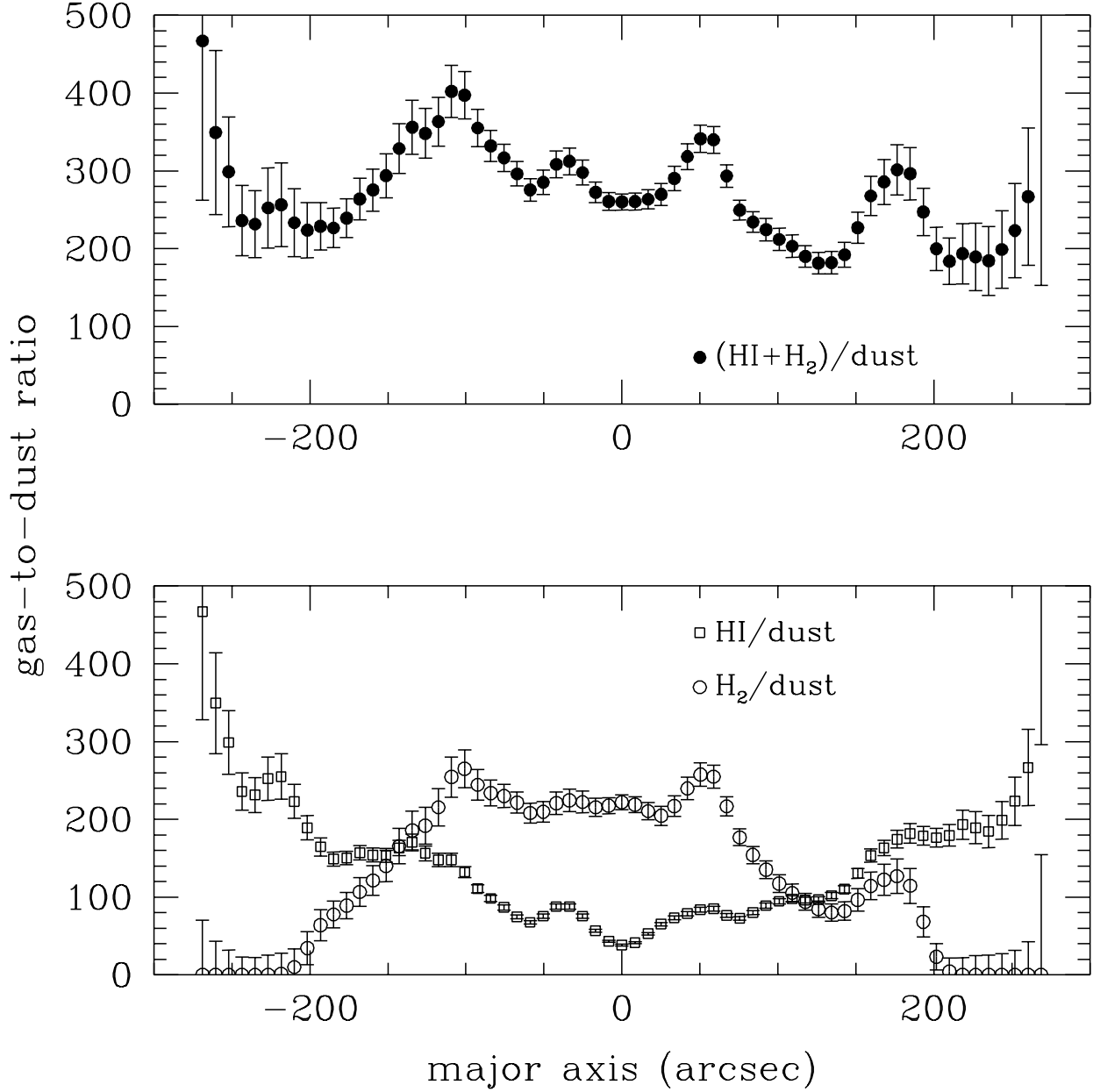}}
\caption{Gas-to-dust mass ratio along the major axis of NGC 891. Solid circles show the ratio of (HI+H$_{2}$)-to-dust. At the bottom, open circles and open squares denote, respectively, the ratio of H$_{2}$-to-dust and HI-to-dust.}
\label{fig5}
\end{figure}

\begin{figure}[hp!]
\centering
\resizebox{!}{21cm}{\includegraphics*[0mm,35mm][210mm,280mm]{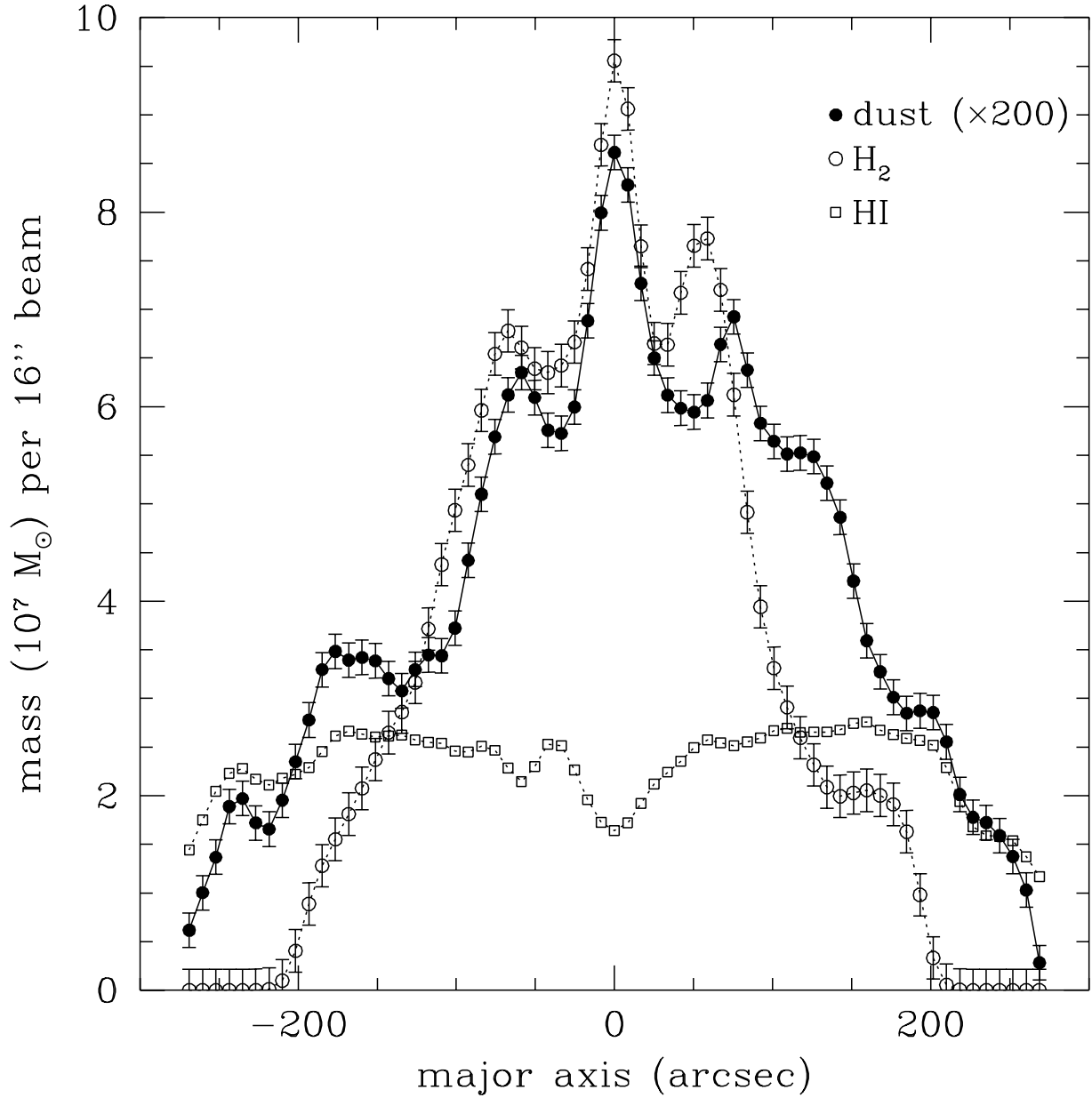}}
\caption{Dust and neutral gas profiles along the major axis of NGC 891. The solid circles represent the distribution of grains (the mass in this case has been multiplied by 200 for the sake of clarity). The open circles and open squares show the distribution of molecular and atomic gas, respectively, within the disk. The poisson error in the HI profile is smaller than the plotted markers.}
\label{fig6}
\end{figure}

In fig.~\ref{fig5}, we plot the gas-to-dust ratio along the major axis of NGC 891. For comparison, we also show the corresponding ratio if either of the neutral gas components (HI or H$_{2}$) is excluded from the calculation. We can see that, at most radii, the molecular hydrogen makes the dominant contribution to the gas mass. Futhermore, the distribution of H$_{2}$ in NGC 891 seems to mimic that of the grain material very closely (see fig.~\ref{fig6}). There are significant fluctuations in the gas-to-dust ratio brought about by an imperfect matching between the submm and CO major axis profiles. Secondary maxima present in both the H$_{2}$ and dust profiles, possibly attributable to spiral arms seen edge-on, are somewhat displaced with respect to each other. `Misalignments' between CO emission and dust extinction lanes in more face-on spirals such as M83 and M100 have already been noted by other observers, although it is still unclear whether this is a more subtle effect of density wave propogation or simply CO acting as a poor tracer of H$_{2}$ (Rand \& Kulkarni 1990; Rand 1995; Rand et al 1999). In the case of NGC 891, fig.~\ref{fig6} seems to indicate that the overall grain distribution is also more radially extended than the layout of molecular gas. Indeed, it is possible that the submm traces the distribution of {\em atomic} gas for radii $r>200''$. Submm observations closer to the edge of the optical disk would be necessary to confirm whether this is a continuing trend since our $850\mu$m data extend to only $r=270''$ whilst 21cm emission is detected out to $r=460''$ ($1.1\times R_{25}$). The gas-to-dust ratio at the `edge' ($\sim R_{25}$) of spiral disks remains something of a mystery. However, background galaxies viewed through such outlying regions appear to be systematical reddened by atomic gas possessing a relatively high gas-to-dust ratio ($\sim 300$; Lequeux and Guelin 1996). Likewise, $200\mu$m imaging of nearby spirals, using ISO, are also indicative of quite large dust scale-lengths, suggesting an association between cold dust and atomic hydrogen at large radii (Alton et al 1998). Our present observations indicate a gas-to-dust ratio $>300$ towards the optical edge of spiral disks. \\

The prominent association between CO emission and submm flux density within NGC 891 (and its implied correspondance between H$_{2}$ and galactic dust), is echoed by recent results from the SCUBA nearby galaxy survey. Dunne et al (1999) have imaged $\sim 100$ objects of typical optical size $1'$ known to possess strong $60\mu$m flux densities ($>$ 5.24 Jy). Notably, they establish a strong correlation between global $850\mu$m and CO emission. Whilst the same trend is observed in both NGC 891 and the nearby, face-on spiral NGC 6946 (Bianchi et al 2000), it is possible that an even stronger correlation with HI+H$_{2}$ will emerge as submm observations become more sensitive.\\

The error-weighted gas-to-dust ratio in NGC 891 is 260 (fig.~\ref{fig5}). The corresponding value for the solar neighborhood is 150, where the dust mass is estimated from either stellar reddening or precipitation of grain elements from the gas phase (Spitzer 1978; Whittet 1992). Sodroski et al (1994) analyzed the large-scale 140 and $240\mu$m from the Galaxy, as detected by COBE, and thereby established a mean gas-to-dust ratio of 160 in the diffuse ISM. Very recently, a gas-to-dust ratio of 200-500 has been determined in several nearby galaxies using either radiation transfer modelling (Xilouris et al 1999; Block et al 1994) or $200\mu$m ISO imaging observations (Alton et al 1998b). Our determination for NGC 891 is consistent with all these values given that sizeable uncertainties exist in the derived dust masses (via the FIR/submm emissivity) and the quantity of molecular gas (via the $X$ parameter).\\

The total dust mass in NGC 891, for that part of the disk covered by our SCUBA maps ($r\leq 225''$), is $1.9 \times 10^{7}$ M$_{\odot}$. This is towards the low end of the range given in ABR ($1.8-4.8 \times 10^{7}$ M$_{\odot}$). 

\section{Metallicity considerations}
\label{metallicity}

In this section, we attempt to establish the heavy element abundance along the disk of NGC 891. A determination of metallicity may cause us to modify the value of $X$, chosen in the previous section, to convert CO line emission to H$_{2}$ column density. Since the gas-to-dust ratio, at least in the inner half of the disk, is dominated by the molecular rather than atomic gas, some justification for our chosen value of $X$ would seem in order. Due to the obscuring effects of the NGC 891 dust lane, we resort to FIR cooling lines from highly ionized metal species (e.g.~OIII $88{\mu}$m) which are expected to be optically thin even for a galaxy viewed edge-on. Accordingly, we have searched the ISO archive for Long Wavelength Spectrometer (LWS) and Short Wavelength Spectrometer (SWS) observations of NGC 891 and, although scientifically-validated spectra of several HII regions were procured, we could not relate the line strengths nor their ratios to any of the diagnostic models currently available in the literature (Statinska 1990; Spinoglio \& Malkan 1992).\\

We then explored more indirect means of inferring the metallicity in NGC 891. Garnet (1998) has shown that there is a clear trend, spanning 4 magnitudes in B-band brightness, between relative oxygen abundance O/H and absolute galaxy luminosity $M_{B}$ (see also Garnett \& Shields 1987). Although extinction effects are severe in NGC 891 (due to its edge-on orientation), the radiative transfer model of XAD may be used to derive the intrinsic blue luminosity of the galaxy. The object can then be located on the O/H vs.~$M_{B}$ plot of Garnet. Following Garnet, we adopt $H_{0}$=50 km/s(Mpc)$^{-1}$ in order to determine the distance modulus and hence absolute B magnitude of NGC 891. Thus, we derive a value of -21.0 for $M_{B}$. On the Garnet plot this corresponds to an oxygen abundance of 12+log($\frac{O}{H}$)=9.2$\pm$0.2 at 1 B-band scale-length from the nucleus (where 0.2 constitutes the dispersion in the metallicity-magnitude `relation'). The Milky Way has an O/H abundance of 9.1 at 1 B-band scale-length from the centre (Shaver et al 1983), suggesting a very similar metallicity between NGC 891 and our own Galaxy.\\

\begin{figure}[hp!]
\centering
\resizebox{!}{21cm}{\includegraphics*[0mm,35mm][210mm,280mm]{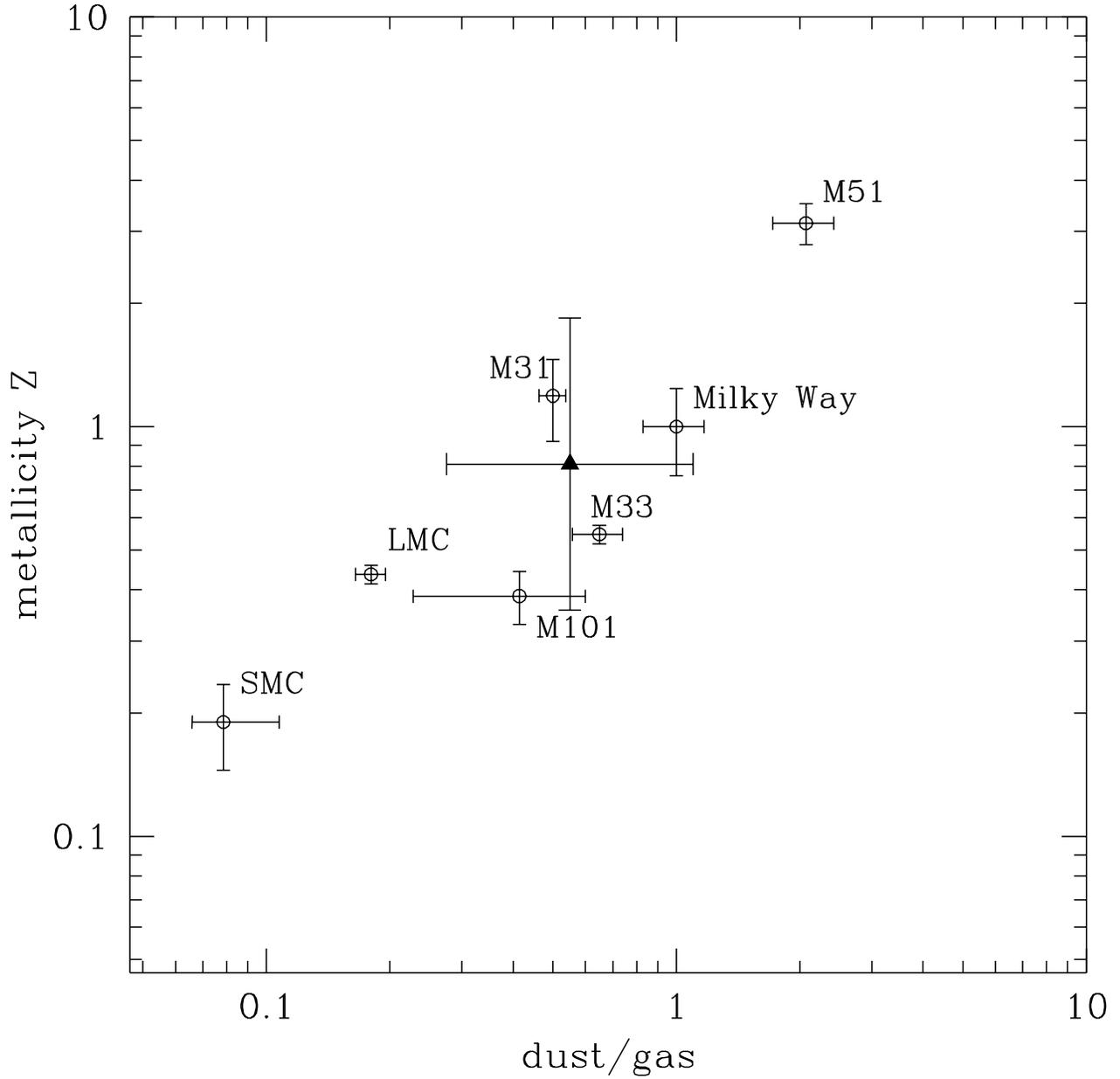}}
\caption{Metallicity against dust-to-gas ratio for NGC 891 compared to galaxies in or near the Local Group (the latter taken from Issa et al 1990). Each quantity is derived at a distance of $0.7R_{25}$ from the centre of the corresponding object and is normalized to the Milky Way. The position of NGC 891 is denoted by a solid triangle and the corresponding errorbar represents principally the uncertainty in the conversion factor $X$ and the dispersion in the metallicity-magnitude relation of Garnet (1998).}
\label{fig6b}
\end{figure}

As mentioned in the previous section, we might expect to modify our value chosen for the conversion factor $X$ if the metallicity in NGC 891 falls below solar (O/H=8.9). For an average metallicity gradient of 0.08 dex/kpc in spiral disks (Vila-Costas \& Edmunds 1992), we expect the oxygen abundance to be 8.9$\pm$0.3 at the maximum radius at which we convert CO line emission to H$_{2}$ column density (r=9.2 kpc). This suggests that the vast majority of molecular gas in NGC 891 has a heavy element abundance similar to, or indeed above, the solar level. Thus the value of $X=2\times 10^{20}$ cm$^{-2}$ K km/s, generally adopted for the general ISM inside the solar ring (\S\ref{gdratio}), can be used with some justification for NGC 891. In fig.~\ref{fig6b}, we reproduce the plot of Issa et al (1990) showing the trend of increasing metallicity  with higher dust-to-gas ratio for galaxies close to or in the Local Group. We incorporate the corresponding quantities derived for NGC 891 which, although subject to appreciable errors, show pleasing agreement with the correlation.

\section{Disk opacity and obscuration effects}
\label{opacity}

Part of the motivation for studying the dust content of external galaxies is to establish whether spiral disks are optically thick or essentially transparent to optical radiation. There exists a whole history of debate concerning the opacity of spiral galaxies. Indeed, various investigative techniques have produced quite disparate answers in the past (Davies \& Burstein 1995; Valentijn 1990; Disney et al 1989). More recent studies seem to be in favour of a V-band optical depth $\sim 1$ through the centre of face-on disks (Xilouris et al 1999; Kuchinski et al 1998), although some authors still maintain spirals are optically thick right out to their $R_{25}$ `edge' (Valentijn 1990). One reason why disk opacity is considered such an important topic is that we view the distant Universe through, what amounts to, a foreground `screen' of spiral galaxies. This has the effect of both attenuating and reddening light emitted by high redshift systems. \\

Alton et al (1999b) have begun to assess the impact of nearby dusty disks on observations of high redshift ($z$) galaxies. For one of the disks which they had observed with ISO (NGC 6946), they adopted a bi-model grain distribution which followed the HI and H$_{2}$ gas components according to a gas-to-dust ratio of 150. This gave an extended dust disk, following the atomic gas, with an exponential fall-off of 14 kpc. Combined with this was a compact grain component, tracing the H$_{2}$, with a radial exponential scale-length of 4.3 kpc. A simulation was run with a Universe populated with disks of this parameterization and assuming a density for field spirals of $\simeq 0.01$ per Mpc$^{3}$ (Loveday et al 1992). Although this constituted a rather simple model (no account was taken of either galaxy clustering or a possible change in gas-to-dust ratio with look-back time) it produced some instructive results. Notably, for an Einstein-de Sitter cosmology and a value of 75 kms$^{-1}$Mpc$^{-1}$ for the Hubble constant, Alton et al (1999b) predicted significant attenuation of light received in the B-band from high redshifts. Indeed, they concluded that 30-40\% of the light originating from $z=2$ would fail to reach the present-day observer due to intervening galactic disks. This attenuation arises principally from the extended dust component which is assumed to follow the atomic hydrogen. The scale-length and, indeed, total extent of this component is far more critical than the central optical depth associated with H$_{2}$ because the peripheral regions of spirals occupy a relatively large filling factor in the plane of the sky. In the absence of information suggesting a radial cut-off for the dust disk, none was applied for this simulation. \\

\begin{figure}[hp!]
\centering
\resizebox{!}{21cm}{\includegraphics*[0mm,35mm][210mm,280mm]{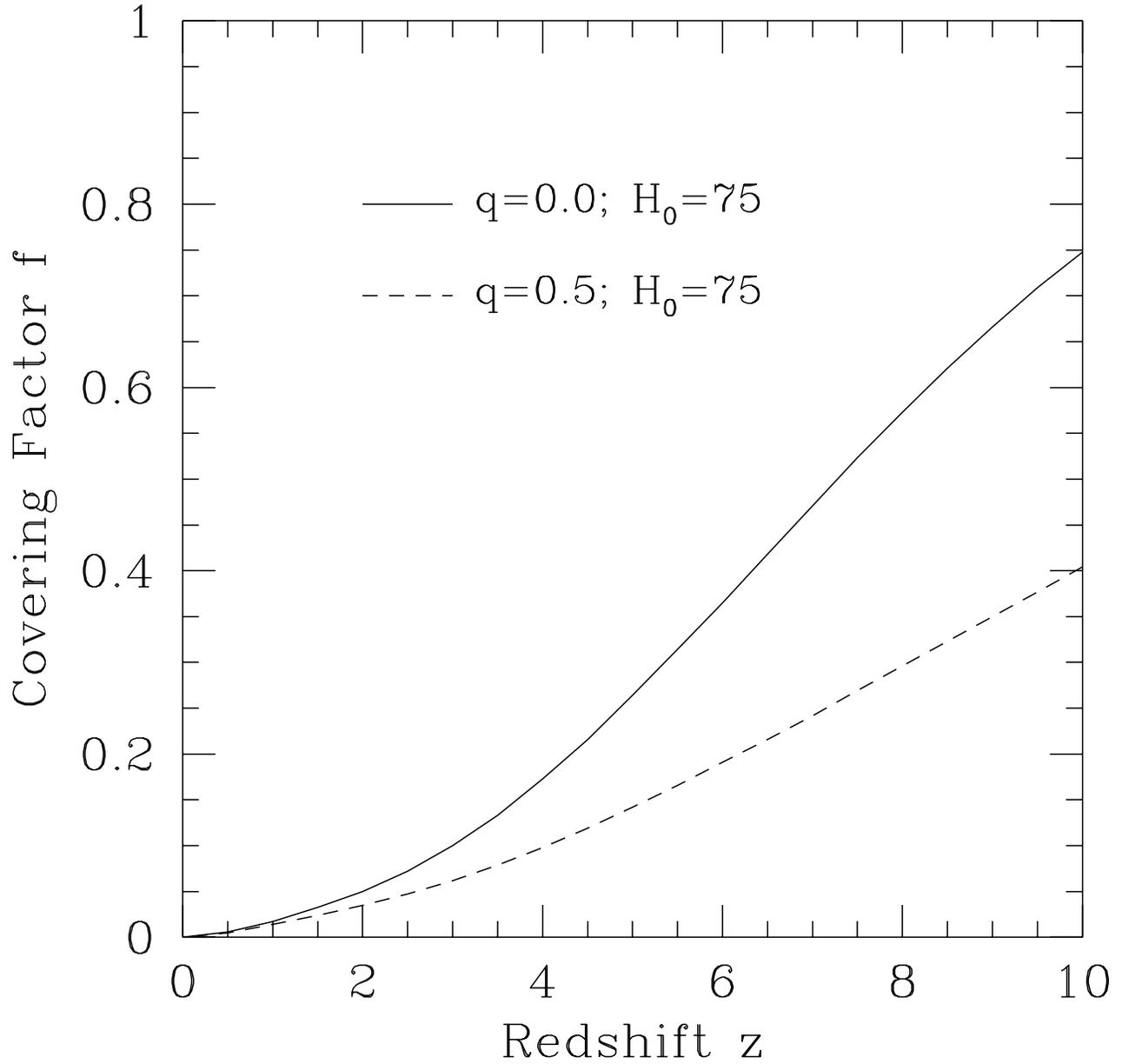}}
\caption{The fraction of light, $f$, emitted at redshift $z$ which fails to reach the B-band observer due to attenuation by foreground spiral disks. We show results for an open Universe and an Einstein-de Sitter cosmology. At all redshifts, a Galactic-type reddening law has been used to calculate the attenuation.}
\label{fig7}
\end{figure}

NGC 6946 is a very gas-rich spiral galaxy (Tacconi \& Young 1990) and, as such, might produce somewhat misleading results when determining high-z visibility (if we simply apply a constant gas-to-dust ratio). Therefore, we wish to repeat the calculations of Alton et al (1999b) using the disk of NGC 891 as a model. We recognize that we will be basing our estimate on a single galaxy but it will be many years before the dust distribution has been accurately mapped out in a large number of nearby galaxies. Thus provisionally, we use the information contained in fig.~\ref{fig5} regarding the gas-to-dust ratio along the major axis of NGC 891. One component of dust is given a radial exponential scale-length of 3.9 kpc, so that it follows the distribution of H$_{2}$ with a gas-to-dust ratio of 260. We also prescribe a second grain component, with an exponential scale-length of 150 kpc. This traces the HI plateau with a gas-to-dust ratio of 350. The maximum radial extent of both components is set at 14 kpc, the point at which the HI column density appears to drop off sharply in the 21cm image of Rupen et al (1991). Our calculation only considers field spirals (using the density given by Loveday et al 1992), but allows for the fact that light received in the B-band, by the observer, has been affected by the UV-FUV part of the extinction curve at higher redshifts. For each slice in redshift, the real density of spiral galaxies is calculated and, accordingly, their contribution to the optical depth at that redshift is determined. Finally, we compute a covering factor, $f$, which is defined as the fraction of light emitted at redshift $z$ which fails to reach a B-band observer due to the sum of intervening disks. This quantity will naturally vary according to our chosen cosmology but we show in fig.~\ref{fig7} our estimates for the important values of $q=0$ and $q=0.5$ (H$_{0}$=75 kms$^{-1}$Mpc$^{-1}$). There is also a small dependance of the solution on the form chosen for the UV reddening law (e.g.~Galactic or LMC-type). However, this is generally neglible for $z<3$. For more detail concerning our computation of $f$ we refer to Trewhella (1997) and Heisler \& Ostriker (1988) which adopt a similar approach to the current method.\\

It is apparent from fig.~\ref{fig7} that, in general, foreground attenuation is not a significant problem for galaxies less than a redshift of 4. Indeed, for galaxies detected in the Hubble Deep Field (assuming typically $z$=1-2 here), less than 5\% of the light appears to be lost. For quasars and related objects, at $z>4$, small corrections are required in order to learn the true luminosity and colour from optical observations. These predictions rely heavily on the cut-off radius of the extended dust component (assumed to follow the HI in the present calculation). When using the NGC 891 SCUBA data, we have extrapolated to $305''$, slightly beyond the furthest radius at which we detect $850\mu$m emission ($270''$). Submm observations at the optical edge of spiral galaxies are required to ascertain the true gas-to-dust ratio and opacity within the outer HI envelope. In contrast, we do not expect the obscuration calculations to be affected severely by our neglect of dust clumping. GMCs have an optical depth of $\tau_{V} \sim 10$ whilst the diffuse ISM is characterized by an optical depth close to unity. Thus, for equal amounts of gas in the molecular and atomic phase, dust clumped into GMCs will occupy a filling factor of only 10\%. As a point of reference, our opacity model for NGC 891, if it were not truncated at $305''$, would give a face-on optical depth of $\tau_{B}=0.25$ at the $R_{25}$ ($408''$).\\

The calculations carried out above take no account of spirals belonging to clusters. At the same time, it is conceivable that a line-of-sight intercepting a foreground cluster may well be subject to significant attenuation. A proper examination of such effects would require inclusion of the correlation function of galaxies and a description of large-scale structure in our computation. To gauge the size of such effects we have calculated the opacity of various lines of sight through the nearby Virgo cluster. Sandage et al (1985) classify 180 cluster spirals of type Sa $\rightarrow$ Sd within a radius of $6^{\circ}$ (2 Mpc) of the Virgo centre. Clearly, there will be a whole range of optical depths through the cluster according to how the light path intercepts the cluster members. Nevertheless, we can evaluate filling factors for regions where $\tau_{V}\ge 1$, $\tau_{V}\ge 0.1$ etc. To do this, we use once again the optical depth inferred from our submm observations of NGC 891 i.e.

\begin{equation}
\label{eq11b}
\tau_{V}(r) = 0.2 e^{-\frac{r}{150}} + 1.9 e^{-\frac{r}{3.9}}
\end{equation}

\noindent for radius $r\leq$14 kpc and $\tau_{V}$=0 for $r>$14 kpc. For $\tau_{V}\geq 1$, the effective cross-section of 180 spiral disks, randomly orientated in the cluster, is 180 $\times$ ($\pi$ 3.5kpc$^{2}$) or 7000 kpc$^{2}$. The total surface area of the Virgo cluster is 4.4$\times$$10^{6}$ kpc$^{2}$, implying that the probability of a line-of-sight possessing an optical depth greater than 1, P($\tau_{V}\geq$1.0), is 0.2\%. At the lower end, P($\tau_{V}\geq$0.2) is limited by the radial cut-off for the dust (14 kpc in eq.~\ref{eq11b}). We derive a probability of only 2\% for a line-of-sight with $\tau_{V}\geq$0.2. This appears to be incompatible with observations of systematic reddening of background galaxies and QSOs by the intergalactic medium (IGM) in nearby clusters. Such studies imply $A_{V}\simeq0.3$ for most line of sights (Boyle et al 1988; Romani \& Maoz 1992). The discrepancy here between calculation and observation would be virtually removed if the radial grain cut-off were relaxed or if significant quantities of dust were believed to be expelled from spiral disks into the IGM during the lifetime of the cluster (Alton et al 1999c; Davies et al 1998). The first of these issues could be addressed by submm observations of cold dust near the optical edge of spiral disks where the HI column density is low ($4\times 10^{20}$ cm$^{-2}$).

\section{Summary}
\label{summary}

We have compared the $850\mu$m emission detected from the nearby, edge-on spiral NGC 891 (Alton et al 1998) with the corresponding optical depth predicted from a sophisticated radiation transfer model (Xilouris et al 1998). Morphologically, both model and submm emission show a very similar distribution along the minor axis, with most of the grain mass contained within a 600pc layer (FWHM). Profiles along the major axis show fair agreement between optical extinction and submm radiation. The $850\mu$m emission is shown to trace the bulk of interstellar dust within NGC 891. Therefore, assuming both extinction and submm emission are attributable to the same grains, we can derive the emissivity (or, equivalently, mass absorption coefficient) at $850\mu$m and, thence, the distribution of dust mass along the disk. We believe that this technique is more robust than taking estimates of FIR emissivity for Galactic grains (Hildebrand 1983; Draine and Lee 1984), and extrapolating them to submm wavelengths. The $850\mu$m emissivity we derive for NGC 891 is $Q(850{\mu}m)\simeq\frac{1}{10600}$, which is about a factor of 2-3 higher than the Draine and Lee (1984) model of Galactic dust. Uncertainty in our technique means that the emissivity in NGC 891 could be a factor of 2 higher or lower than the quoted value. Importantly, given that the employed radiation transfer model neglects dust-clumping, a {\em decrease} of factor 2 seems quite likely.\\

We have computed the gas-to-dust ratio along the major axis of NGC 891 using our newly-acquired submm emissivity. We derive a value of 260, which is consistent, within the uncertainties, with  estimates for the diffuse ISM of the Milky Way and nearby spirals. The dust in NGC 891 appears to be closely associated with the molecular gas phase, at least for the inner half of the disk. At radii $>$9 kpc ($> \frac{1}{2} R_{25}$), it is possible the grains begin to follow the distribution of atomic hydrogen. We use the opacity of the NGC 891 disk to calculate the degree with which light emitted at high redshift is attenuated by foreground field spirals. For galaxies typically associated with the Hubble Deep Field (z=1-2), the fraction of light lost, in this way, is only 5\%. Since this estimate is based on a single galaxy as a model, further submm measurements will be required to support our conclusions. In particular, observations at the periphery of spiral disks could establish the extent to which cold dust is associated with the expansive HI envelopes surrounding spiral galaxies (it is the maximum radial extent of the dust which primarily controls the level of foreground extinction). Futhermore, we point out that certain lines-of-sight through nearby clusters are likely to be much more optically thick (than our figure of 5\% would suggest), particularly if grains are expelled into the intergalactic medium via starburst winds (Alton et al 1999c) or dispersed through tidal interactions (Yun et al 1994).\\

Finally, we note the strong correspondance between dust and molecular gas in NGC 891. Further submm imaging of face-on spirals (both H$_{2}$ rich and H$_{2}$ poor) would confirm the trend indicated by the current observations, that the bulk of galactic dust appears to be associated with the molecular gas phase.


\begin{acknowledgements}
We would like to thank Loretta Dunne and Steve Eales for revealing early results from their SCUBA nearby galaxy survey. It is also a pleasure to thank Adolf Witt for his comments on dust clumping in spiral disks. We acknowledge the assistance of Sunil Sidher in the reduction of ISO LWS spectra. As always, PBA acknowledges PPARC for its continued financial support. 
\end{acknowledgements}

\section*{References}

\noindent Agladze, N., Sievers, A., Jones, S., Burlitch, J., Beckwith, S., 1994, Nature, 372, 243

\noindent Alton, P., 1996, Ph.D.~University of Durham

\noindent Alton, P., Davies, J., Trewhella, M., 1998a, MNRAS, 296, 773

\noindent Alton, P.B., Trewhella, M., Davies, J., Evans, R., Bianchi, S, Gear, W., Thronson, H., Valentijn, E., Witt, A., 1998b, A\&A 335, 807

\noindent Alton, P., Bianchi, S., Rand, R., Xilouris, E., Davies, J., Trewhella, M., 1998c, ApJL, 507, L125 (ABR)

\noindent Alton, P., Rand, R., Xilouris, E., Davies, J., Bianchi, S., 1999a, A\&A, submitted

\noindent Alton, P., Bianchi, S., Davies, J., 1999b, ApSS in press

\noindent Alton, P., Davies, J., Bianchi, S., 1999c, A\&A, 343, 51

\noindent Bianchi, S., Alton, P., Davies, J., Trewhella, M., 1998, MNRAS, 298, L49

\noindent Bianchi, S., Davies, J., Alton, P., 1999a, A\&A, 344, L1

\noindent Bianchi, S., Ferrara, A., Davies, J., Alton, P., 1999b, MNRAS in press

\noindent Bianchi, S., Davies, J., Alton, P., Gerin, M., Casoli, F., 2000, A\&A, 353, L13

\noindent Block, D., Witt, A., Grosbol, P., Stockton, A., Moneti., A., 1994, A\&A, 288, 383

\noindent Bohlin, R., Savage, B., Drake, J., 1978, ApJ, 224, 132 

\noindent Bohren, C., Huffman, D., 1983, in {\em Absorption \& Scattering of Light by Small Particles} (John Wiley \& Sons)

\noindent Boyle, B., Fong, R., Shanks, T., 1988, MNRAS, 231, 897

\noindent Cardelli, J., Clayton, G., 1988, AJ, 95, 516

\noindent Casey,S., 1991, ApJ, 371, 183

\noindent Chini, R., Kruegel, E., Lemke, R., Ward-Thompson, D., 1995, A\&A, 295, 317

\noindent Clements, D., Andreani, P., Chase, S., 1993, MNRAS, 261, 299

\noindent Davies, J.I. \& Burstein, D., 1995, in `The opacity of Spiral Disks', proceedings of NATO ARW, NATO Asi Ser.~469, ed.~J.Davies and D.Burstein

\noindent Davies, J., Alton, P., Bianchi, S., Trewhella, M., 1998, MNRAS, 300, 1006 

\noindent Desert, F., Boulanger, F., Puget, J., 1990, A\&A, 237, 215

\noindent Dettmar, R-J., 1990, A\&A, 232, L15

\noindent Devereux, N.A. \& Young, J.S., 1990, ApJ, 359, 42

\noindent Disney, M.J., Davies, J.I. \& Phillips, S., 1989, MNRAS, 239, 939

\noindent Draine, B., Lee, H., 1984, ApJ, 285, 89

\noindent Dunne, L., Eales, S., Edmunds, E., Ivison, R., Alexander, P., Clements, D.,  1999, MNRAS, in press

\noindent Garcia-Burillo, S., Guelin, M., Cernicharo, J., Dahlem, M., 1992, A\&A, 266, 21

\noindent Garnet, D., Shields, G., 1987, ApJ, 317, 82

\noindent Garnet, D., 1998, in {\em Abundance Profiles: Diagnostic Tools for Galaxy History} ed Friedli, Edmunds, Robert, Drissen

\noindent Guelin, M, Zylka, R., Mezger, P., Haslam, C., Kreysa, E., Lemke, R.,  Sievers, A., 1993, A\&A 279, L37

\noindent Heisler, J., Ostriker, J., 1988, ApJ, 332, 543

\noindent Henyey, L., Greenstein, J., 1941, ApJ, 93, 70

\noindent Hildebrand, 1983, QJRAS, 24, 267

\noindent Holland, W., Robson, E, Gear, W., et al, 1999, MNRAS, 303, 659

\noindent Howk, J., Savage, B., 1997, AJ, 114, 2463 

\noindent Howk, J., Savage, B., 1999, AJ, 117, 2077

\noindent Hughes, D., Robson, E., Dunlop, J., Gear, W., 1993, MNRAS, 263, 607

\noindent Hughes, D., Dunlop, J., Rawlings, S., 1997, MNRAS, 289, 766

\noindent Israel, F., van der Werf, P., Tilanus, R., 1999, A\&A, 344, L83

\noindent Issa, M., MacLaren, I., Wolfendale, A., 1990, A\&A, 236, 237

\noindent Kuchinski, L., Terndrup, D., Gordon, K., Witt, A., 1998, AJ, 115, 1438

\noindent Kylafis, K., Xilouris, M., et al, 1999, ApSS, in press

\noindent Larson, R., 1981, MNRAS, 194, 809

\noindent Lequeux, J., Guelin, M., 1996, in {\em New Extragalactic Perspectives in the new South Africa}, (Kluwer Academic Publishers)

\noindent Loveday, J., Peterson, B., Efstathiou, G., Maddox, S., 1992, ApJ, 390, 338

\noindent Maloney 1990, in {\em Interstellar Medium of Galaxies} ed.~Thronson, H., Shull, J.

\noindent Masi, S, Aquilini, E., Boscaleri, A., de Bernardis, P., de Petris, M., Gervasi, M., Martinis, L., Natale, V., Palumbo, P., Scaramuzzi, F., 1995, ApJ, 452, 253

\noindent Mathis, J., Whiffen, G., 1989, ApJ, 341, 808

\noindent Mathis, 1990, ARAA, 28, 37

\noindent Mathis, J., 1997, {\em From Stardust to Planetesimals}, ASP Conference Series, Vol.~122, edt.~Pendleton \& Tielens

\noindent Neininger, N., Guelin, M., Garcia-Burillo, S., Zylka, R., Wielebinski, R., 1996, A\&A, 310, 725

\noindent Pendleton, Y, Tielens, A., Werner, M., 1990, ApJ, 349, 107

\noindent Polk, K., Knapp, G., Stark, A., Wilson, R., 1988, ApJ, 332, 432

\noindent Rand, R., Kulkarni, S., 1990, ApJ, 349, L43

\noindent Rand, R., 1995, AJ, 109, 2444

\noindent Rand, R., Lord, S., Higdon, J., 1999, ApJ, 513, 720

\noindent Reach, W., Dwek, E., Fixsen, D. et al, 1995, ApJ, 451, 188

\noindent Romani, R., Maoz, D., 1992, ApJ, 386, 36

\noindent Rupen, M., 1991, AJ, 102, 48

\noindent Sandage, A., Binggeli, B., Tammann, G., 1985, AJ, 90, 395 

\noindent Scoville, N., Thakkar, D., Carlstrom, J., Sargent, A., 1993, ApJ, 404, L59

\noindent Shaver, P., McGee, R., Newton, L., Danks, A., Pottasch, S., 1983, MNRAS, 204, 53

\noindent Sodroski, T., et al, 1994, ApJ, 428, 638

\noindent Spinoglio, L., Malkan, M., ApJ, 399, 504

\noindent Spitzer, 1978, {\em Physical Processes in the Interstellar Medium} (John Wiley \& Sons)

\noindent Stasinska, G., 1990, A\&ASS, 83, 510

\noindent Sukumar, S., Allen, R., 1991, ApJ, 382, 100

\noindent Tacconi, L., Young, J., 1990, ApJ, 253, 595

\noindent Trewhella, M., (1997) Ph.D. University of Wales, Cardiff

\noindent Valentijn, E., 1990, Nature, 346, 153

\noindent Vila-Costas, M., Edmunds, M., 1992, MNRAS, 259, 121

\noindent Whittet, D., 1992, {\em Dust in the Galactic Environment} (IOP Publishing)

\noindent Witt, A., Gordon, K., Madsen, G., 1999, ApJ, submitted

\noindent Xilouris, E., Kylafis, N, Papamastorakis, J., Paleologou, E., Haerendel, G., 1997, A\&A, 325, 135 

\noindent Xilouris, E., Alton, P., Davies, J., Kylafis, N., Papamastorakis, J., Trewhella, M., 1998, A\&A, 331, 894 (XAD)

\noindent Xilouris, E., Byun, Y., Kylafis, N., Paleologou, E., Papamastorakis, J., 1999, A\&A, 344, 868

\noindent Young, J., Scoville, N., 1991, ARAA, 29, 581

\noindent Yun, M., Ho, P., Lo, K., 1994, Nature, 372, 530

\newpage

\begin{table}[hp!]
\centering
\caption{Properties of the grain and stellar disks in NGC 891. The Right Ascension (R.A.) and Declination (Dec.) are given by the B1950.0 nuclear position recorded by Sukumar \& Allen (1991). The remaining parameters are derived from a V-band radiation transfer simulation carried out by Xilouris et al (1998). $z_{S}$ and $z_{D}$ denote the exponential scale-height of the stars and dust respectively. Similarly, $h_{S}$ and $h_{D}$ represent the exponential scale-length of the stellar and dust disks. $\tau^{f}$ is the V-band optical depth through the centre of the disk, if the galaxy were to be viewed exactly face-on. $\theta$ is the inclination of the disk with respect to the plane of the sky.}
\label{tab1}
\begin{tabular}{cc}

Parameter  & Value \\ \\
 R.A.	&  2h 19 24.3 \\
 Dec.   &   42$^{\circ}$ 07 17 \\
 $z_{S}$ (kpc)    		&  0.39 \\
 $z_{D}$ (kpc)   		&  0.26 \\
 $h_{S}$ (kpc)    		&  5.71 \\
 $h_{D}$ (kpc)    		&  8.10 \\
 $\tau^{f}$			& 0.70 \\
 $\theta$     			&  89.8 \\
\\

\end{tabular}
\end{table}

\newpage

\end{document}